\documentclass[twocolumn,prl,superscriptaddress,noshowpacs]{revtex4}
\usepackage{graphicx}

\begin{document}
\title{Charge fluctuation induced dephasing of exchange coupled spin qubits}
\date{\today}
\author{Xuedong Hu}
\affiliation{Department of Physics, University at Buffalo, The State
University of New York, 239 Fronczak Hall, Buffalo, NY 14260-1500}
\author{S. Das Sarma}
\affiliation{Condensed Matter Theory Center, Department of Physics, University
of Maryland, College Park, MD 20740-4111}
\begin{abstract}
Exchange coupled {\it spin} qubits in semiconductor nanostructures are
shown to be vulnerable to dephasing caused by {\it charge noise}
invariably present in the semiconductor environment.  This decoherence of
exchange gate by environmental charge fluctuations arises from the
fundamental Coulombic nature of the Heisenberg coupling, and presents a
serious challenge to the scalability of the widely studied exchange gate
solid state spin quantum computer architectures.  We estimate dephasing
times for coupled spin qubits in a wide range (from 1 ns up to $> 1 \mu$s)
depending on the exchange coupling strength and its sensitivity to charge
fluctuations.
\end{abstract}
\maketitle

A central issue in quantum information processing is quantum coherence, i.e.
how long a quantum state survives without decay allowing robust quantum
computation.  The other central issue is scalability, i.e. whether a
practical {\it macroscopic} quantum computer can be built by suitably
scaling up individual {\it microscopic} qubits.  The perverse dichotomy in
quantum computation has been that architectures that can be scaled up fairly
easily (e.g. solid state systems) suffer from serious environmental
decoherence problems whereas architectures based on isolated atoms and ions,
which have excellent coherence (e.g. ion traps), are typically not easily
scalable.  In this context, the proposed spin quantum computer (QC)
architectures
\cite{LD,Kane,DiVincenzo,HD,SQCReviews} in semiconductor nanostructures look
particularly promising since electron spin usually has long coherence time as
the relativistic nature of spin produces weak direct environmental
coupling, and semiconductors, at least as a matter of principle, allow for
relatively easy scaling up.  Motivated by the pioneering early suggestions
\cite{LD,Kane,DiVincenzo}, there has been impressive recent experimental
advance in the study of spin qubits in gated GaAs quantum dot systems
\cite{Elzerman,Johnson,Hatano,Petta}, an architecture widely regarded as one
of the most promising solid state QC architectures.

One of the most significant advantages of a spin qubit is its relative
isolation from its environment, leading to exceedingly long relaxation
and dephasing times in systems such as isolated donor electron and
nuclear spins in Si:P and GaAs quantum dots \cite{Feher,Fujisawa}. 
Specifically, electron spin relaxation due to spin-orbit interaction and
coupling to phonons is significantly reduced at low temperatures and
for localized spins \cite{Golovach}.  The only important decoherence channel
left is the nuclear spin hyperfine coupling induced electron spin spectral
diffusion, so that the spin decoherence time for an isolated electron could
range from tens of microsecond in GaAs quantum dots up to hundreds of
millisecond in Si:P donor electrons \cite{Rogerio}.

New decoherence channels could open up when qubits are manipulated and/or
coupled.  An advantage of spin qubits in solids is the
availability of exchange interaction, which originates from Coulomb
interaction and Pauli exclusion principle.  Even though spins are magnetic
and magnetic interactions are weak, two-spin operations can actually be very
fast (as short as 100 ps), because the underlying exchange interaction is
electrostatic, which is strong.  However, therein lies a new decoherence
channel (compared to single spins) for the spin qubits: when spin
coupling is needed and exchange interaction is turned on using external
gates, charge fluctuations in the environment (an important source of
decoherence for charge qubits in semiconductor \cite{Hayashi} and 
superconducting \cite{Astafiev} structures) could lead to gate errors and
dephasing \cite{Burkard}.   

In this Letter we quantify how charge fluctuations affect exchange gates for
spin qubits in a double quantum dot.  In particular, we calculate the
modification of exchange coupling in the presence of barrier variation and
double dot level detuning (both of which could arise from charge fluctuations
in the environment), evaluate errors in exchange gates, and calculate
dephasing rates for logical qubits encoded in the double dot
singlet-triplet states.

\begin{figure}[t]
\begin{center}
\includegraphics[width=3.4in]{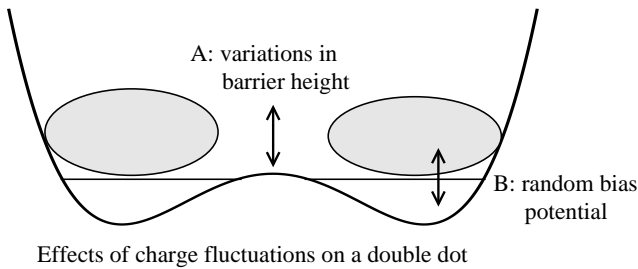}
\caption[How does charge movement affect DQDs]{
Charge fluctuations invariably present in the semiconductor environment affect
a double quantum dot mainly by causing variations of barrier potential height
(A) and producing a random bias potential between the two dots (B).  
}
\label{fig1}
\end{center}
\end{figure}

Gated quantum dots are defined electrostatically from a two-dimensional
electron gas (2DEG).  A movement of a trapped charge closeby
changes the confining potential mainly in two ways: rise or fall of the
barrier between the dots (thus change in tunneling rates), and detuning of
the orbital levels in the two dots \cite{Jung}, as illustrated in
Fig.~\ref{fig1}.  When the central barrier between the double dot rises or
falls, the dot potential minima also shift slightly farther or closer from
each other.  

With this qualitative understanding of how charge fluctuations affect
quantum dot confinement, we evaluate the exchange coupling of two electrons in
a gate-defined unbiased double quantum dot in the presence of charge
fluctuations.  We use a two-dimensional quartic potential
\cite{growth_direction,Burkard}:
$V(x,y) = \frac{1}{2} m \omega^2 [(x^2-L^2)^2/L^2 + y^2]$.
The central barrier height for this potential is $V_B = m\omega^2 L^2/2$,
directly related to the interdot distance \cite{footnote-Barrier}.  We use
GaAs parameters in this calculation,
with $m = 0.067 m_0$ where $m_0$ is the bare electron mass.  To calculate the
exchange splitting for two electrons in this double dot potential, we employ
the Heitler-London (HL) approximation \cite{Herring}, so that the exchange
splitting $J$ between the two-electron singlet and unpolarized triplet states
can be expressed in terms of a few easy-to-calculate matrix elements:
\begin{eqnarray}
J & = & \frac{2S^2}{1-S^4} \left[ \langle \psi_L | V-V_L | \psi_L \rangle +
\langle \psi_R | V-V_R | \psi_R \rangle \right] \nonumber \\
& & + \frac{2S^2}{1-S^4} \langle
\psi_L(1) \psi_R(2)| \frac{e^2}{\epsilon r_{12}} |\psi_L(1) \psi_R(2) \rangle
\nonumber \\
& & -\frac{2S}{1-S^4} \left[ \langle \psi_L | V-V_R | \psi_R \rangle + \langle
\psi_R | V-V_L | \psi_L \rangle \right] \nonumber \\
& & - \frac{2}{1-S^4} \langle \psi_L(1)
\psi_R(2)| \frac{e^2}{\epsilon r_{12}} |\psi_L(2) \psi_R(1) \rangle\,.
\label{eq:exchange}
\end{eqnarray}
Here $V_L$ and $V_R$ are harmonic potential wells that share the same location
as the left and right potential minima of given potential $V$, $\psi_L$
or $\psi_R$ is the ground state if only $V_L$ or $V_R$ is present, indices 1
and 2 refer to the two electrons, $S$ is the overlap integral between the
single dot orbitals, and $r_{12}$ is the inter-electron distance.  The first
and third terms in Eq.~(\ref{eq:exchange}) refer to single particle
contributions to $J$, the second term originates from the
direct repulsion between the two electrons, and the last term is the exchange
contribution.
The HL approximation works well for higher confinement energies ($\hbar \omega
\gtrsim 3$ meV, corresponding to smaller dots) when the on-site Coulomb
interaction is sufficiently large, and for larger interdot distances.  
\begin{figure}
\includegraphics[width=3.4in]
{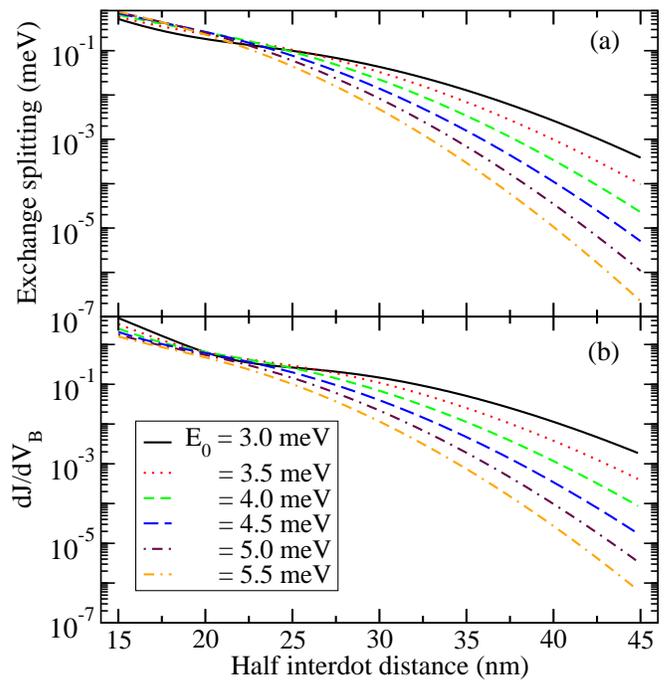}
\protect\caption[Exchange and its dependence on V]
{\sloppy{
Panel (a) presents exchange splitting as a function of inter-dot distance
(directly related to the inter-dot barrier height) for various dot sizes
(directly related to the single dot confinement energy $E_0$) with the quartic
double dot
potential, while panel (b) presents $dJ/dV_B$ (derivative of $J$ over
inter-dot barrier potential) as a function of
inter-dot distance.  Panel (b) gives a quantitative estimate of the
sensitivity of exchange coupling to background charge fluctuations.
}}
\label{fig2}
\end{figure}

In Fig.~\ref{fig2} we plot the exchange splitting and the
barrier-height-$V_B$-dependence of the exchange for a series of
configurations of double dots.  For strongly coupled dots exchange splitting
up to 1
meV can be obtained, where $dJ/dV_B$ can be larger than 1, rendering the
coupled spin qubits as sensitive to charge noise as a charge qubit.  Notice
that at the large exchange limit, more
sophisticated method is required for reliable evaluation of the
exchange splitting \cite{HD,Burkard}, though the modifications are generally
within one order of magnitude so that the qualitative feature of our current
results remains.  Charge fluctuations could also introduce a small
bias $\Delta V$ between the originally unbiased double dot, which leads to a
second-order correction [$\Delta J \propto (\Delta V)^2$].  For relatively
small charge fluctuations, we neglect this bias
effect on exchange since it is a higher-order effect.

In some experimental situations the double dot is strongly biased so that one
of the doubly-occupied two-electron singlet states is the ground state
\cite{Johnson,Petta}.  In this case the exchange splitting between the
two-dot two-electron singlet and triplet states could be dominated by the
tunnel coupling between the two- and one-dot singlet states
(while the triplet cannot tunnel due to spin blockade \cite{Ono}), and
takes on the value $J \approx |t|^2/E_b$, where $t$ is the tunnel coupling
between the two singlet states and $E_b$ their energy difference, 
dominated by the applied bias potential between the two dots.  Charge
fluctuations affect both $t$ and $E_b$, so that $dJ = (\frac{2J}{t}
\frac{dt}{dV_B}) dV_B - (\frac{J}{E_b}) dE_b$ (assuming $dV_B$ and $dE_b$
represent different components of charge noise), where $V_B$ is the barrier
potential that determines tunnel coupling $t$.  If $dJ$ is dominated by
$dE_b$, $dJ = -(J/E_b) dE_b$.  For $J \sim 1 \mu$eV and $E_b \sim 100 \mu$eV,
$|dJ/dE_b| \sim 0.01$, similar to a weakly coupled unbiased
situation. (Here $dJ/dE_b$, as opposed to $dJ/dV_B$, controls the charge
fluctuation effect on the exchange gate.)  Our estimates here are
consistent with the experimentally measured exchange dependence on bias
voltage presented in Ref.~\cite{Petta}, where $dJ/dE_b$ ranges
between 0.01 and $10^{-4}$, and falls in
the same range as our theoretical results in Fig.~\ref{fig2}. 

Having analyzed how charge fluctuations affect the exchange
splitting of a double dot, we now study how spin
quantum computing could be influenced.  Charge fluctuations can affect
coupled spin qubits in two different ways.  If exchange interaction is turned
on briefly to perform two-qubit operations, a switching event of a nearby
charge trap leads to a gate error. 
If the exchange interaction is
constantly on and the singlet and the unpolarized triplet states are the
logical qubit states, background charge fluctuation causes pure
dephasing between the two states.  Below we analyze these two situations
separately.

A switching event during an exchange gate inevitably leads to a gate
error.  Consider a SWAP gate, where an exchange pulse corresponding to
$\int J dt/\hbar = \pi$ leads to a swap of the states of two spins.  If
instead $\int J dt/\hbar = \pi + \delta$, the result of the exchange pulse is
\begin{eqnarray}
\left(\begin{array}{c}
\alpha_1 \\
\alpha_2
\end{array}
\right) 
\left(\begin{array}{c}
\beta_1 \\
\beta_2
\end{array}
\right) & \stackrel{\rm SWAP}{\longrightarrow} &
\left(\begin{array}{c}
\beta_1 \\
\beta_2 
\end{array}
\right) 
\left(\begin{array}{c}
\alpha_1 \\
\alpha_2 
\end{array}
\right) \nonumber \\ 
& & + \frac{\alpha_1 \beta_2 - \alpha_2 \beta_1}{\sqrt{2}} \left(
1-e^{i\delta} \right) 
|S\rangle \,.
\end{eqnarray}
Here $|S\rangle = (|\uparrow \downarrow\rangle - |\downarrow \uparrow\rangle)
/\sqrt{2}$ is the two-spin singlet state.  Thus the two spins would
remain entangled after the SWAP operation.
This error is linearly proportional to $\delta$ when $\delta \ll \pi$, 
and its prefactor $\alpha_1 \beta_2 - \alpha_2 \beta_1$ does not vanish unless
the two single spin states are identical. 

If we have three spins and intend to swap the state of the first to the third,
errors accumulate linearly, with residual entanglement between qubits 1 and 3,
and 2 and 3.  The total error should grow with the number of spins $N$
linearly even if errors in each swap are random, since different errors are
not directly additive as they represent different unwanted entanglement after
the swaps.  One can estimate errors in more complicated operations such as a
Controlled-NOT gate \cite{Thorwart}, where the feature of linear increase of
error with the number of qubits persists.

If two-spin singlet and unpolarized triplet states are used as logical qubit
states and the exchange coupling is kept on, background charge fluctuations
cause dephasing between the two states.  Since charge
fluctuation is also a major source of decoherence for double dot charge qubits
\cite{Hayashi}, we can extract the necessary information
needed to calculate spin decoherence rate from the measured charge relaxation
rates in the GaAs quantum dot system.  Similar to what is well established in
the decoherence properties of Cooper pair boxes \cite{Astafiev}, 
the relaxation rate at the degeneracy
point of a double dot single-electron charge qubit (where the double dot
ground and first
excited states are split by $2|t|$ where $t$ is the tunneling
strength between the two dots.  The Hamiltonian for such a charge qubit can be
written as $H=|t|\sigma_z + V\sigma_x$ where $V$ is the potential bias
between the two dots) is given by a simple expression \cite{Astafiev}
$\Gamma_1 = (\pi/2\hbar^2) S_V (\omega = 2|t|/\hbar)$, where $S_V (\omega) =
(1/2\pi) \int_{-\infty}^{\infty} d\tau e^{i\omega\tau} \langle V(\tau) V(0)
\rangle$ is the charge fluctuation (in terms of the bias gate potential
fluctuation) correlation in the environment.  From the functional dependence
of $S_V(\omega)$ on $\omega$ (such as $1/f$ noise, which has only one
parameter and is used here),
we could use the knowledge of $\Gamma_1$ to determine $S_V(\omega)$, and then
use it to calculate the time($\tau$)-dependent phase diffusion $\Delta
\phi_c$ for a charge qubit (so that a factor $\exp (-\Delta \phi)$ appears in
the
off-diagonal density matrix element of the two-level system formed from the
double dot) when it is far away from the degeneracy point, where the effective
Hamiltonian becomes $H=V\sigma_z$:
\begin{equation}
\Delta \phi_c (\tau) = \frac{1}{2\hbar^2} \int_{\omega_0}^{+\infty} d\omega
S_V(\omega) \left( \frac{\sin \omega \tau/2}{\omega/2} \right)^2 \,.
\label{eq:charge_dephasing}
\end{equation}
Here the integral has a low frequency cutoff that is generally taken as the
inverse of the measurement time \cite{Astafiev}.
In the case of two-spin singlet and unpolarized
triplet states, the effective two-level Hamiltonian can be written as
$H=J\sigma_z$ (there is no $\sigma_x$ term here as spin symmetry prevents
direct relaxation between the two states), so that the two-spin dephasing is
given by \cite{Astafiev}
\begin{eqnarray}
\Delta \phi_s (\tau) & = & \frac{1}{2\hbar^2} \int_{\omega_0}^{+\infty}
d\omega S_J(\omega) \left( \frac{\sin \omega \tau/2}{\omega/2} \right)^2
\nonumber \\
& \cong & \left(\frac{dJ}{dV}\right)^2 \Delta \phi_c (\tau) \,,
\label{eq:spin_dephasing}
\end{eqnarray}
where $(dJ/dV)^2 = (\partial J/\partial V_B)^2 + (\partial J/\partial E_b)^2$. 
Accordingly, two-spin dephasing should be sensitively dependent
on the barrier- and bias-dependence of the exchange splitting $(dJ/dV)^2$. 
Equation~(\ref{eq:spin_dephasing}) is valid when $S_{V_B}(\omega)$ for the
inter-dot barrier height and $S_{V=E_b}(\omega)$ for the inter-dot bias are
identical (recall that they originate from the same source of charge
fluctuations and therefore should have similar behavior).  

\begin{figure}[t]
\begin{center}
\includegraphics[width=3.4in]{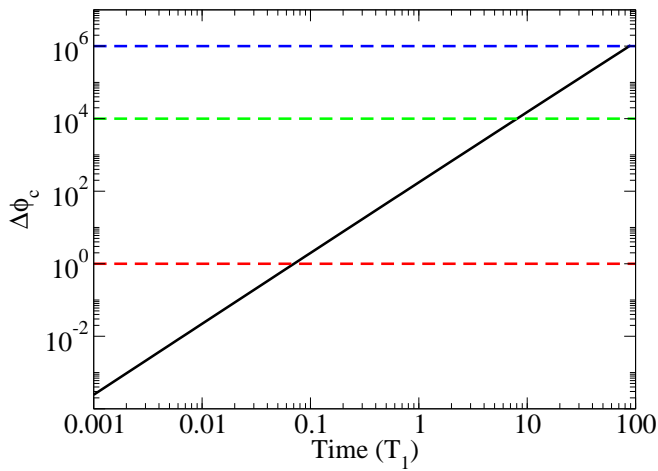}
\caption[Phase diffusion due to charge fluctuations]{
Phase diffusion of a biased double dot charge qubit with a cutoff
frequency of 1 Hz.  Dephasing of a two-spin
qubit can be determined from this figure as well since it is linearly
proportional to the charge qubit phase diffusion, as illustrated by
Eq.~(\ref{eq:spin_dephasing}).  The lower horizontal line indicates that
$\Delta \phi_c \sim 1$ corresponds to $t \sim T_1/10$, the middle (upper)
horizontal line indicates that $\Delta \phi_c \sim 10^4$ ($10^6$) corresponds
to $t \sim 10 T_1$ ($100 T_1$).  In other words, if $dJ/dV \sim 0.001$,
$\Delta \phi_s \sim 1$ would corresponds to a time of $100 T_1$ where charge
relaxation time $T_1$ is in the order of 10 ns, so that spin dephasing time
$\sim 1 \mu$s.
}
\label{fig3}
\end{center}
\end{figure}

Figure~\ref{fig3} shows the charge qubit phase diffusion.  Two-spin phase
diffusion can be obtained according to
Eq.~(\ref{eq:spin_dephasing}).  The phase diffusion grows as
almost a quadratic function of time (actually $t^2\ln t$).  For charge
relaxation time in the order of 10 ns \cite{Hayashi}, charge dephasing time
for biased qubit is in the order of 1 ns (lower horizontal line in
Fig.~\ref{fig3}).  Two-spin
dephasing time sensitively depends on $dJ/dV$, which in turn varies widely
(see Fig.~\ref{fig2} and \cite{Petta}).  For example, if $dJ/dV = 1$
(when the double dot is tightly coupled), the two-spin dephasing time is as
short as the charge dephasing time.  On the other hand, if $dJ/dV = 0.01$,
which is a reasonable estimate (see Fig.~\ref{fig2}), two-spin dephasing time
would be in the order of ten times of the charge relaxation time $T_1
\sim 10$ ns, leading to $T_2 \sim 0.1 \mu$s.  If $dJ/dV$ could be reduced to
0.001, for small $J$ \cite{Petta}, or using schemes that can minimize
$dJ/dV$ through device design, in the spirit of the pseudo-digital design (to
combat gate voltage fluctuations) of Ref.~\cite{Digital}, in which $J$
reaches maximum while $dJ/dV$ is close to zero, the two-spin dephasing time
would be $\sim 1 \mu$s, in the same order as the single spin decoherence time
due to nuclear spin induced spectral diffusion in GaAs \cite{Rogerio}.  In
general, extremely small values of $dJ/dV$ is needed to essentially eliminate
the charge fluctuation induced spin decoherence.

Charge fluctuation induced spin qubit dephasing is obviously not just limited
to GaAs quantum dots.  Whenever charge degrees of freedom (e.g. electrostatic
gates) are used to boost the speed of a quantum computing scheme, dephasing
effect from charge noise in the underlying structure could arise.  For
instance, charge noise could have negative effects on trapped ions when
semiconductor micro-traps are used, so that charge fluctuations in the
semiconductor environment could lead to decoherence in the ionic states. 
Similarly, various proposed solid state quantum computing schemes using only
exchange gate architectures are potentially susceptible to charge noise
decoherence.

In conclusion, although single spin coherence is unaffected by charge noise,
background charge fluctuations could lead to significant gate errors and/or
decoherence in semiconductor-based electron spin qubits through inter-qubit
exchange coupling.  Our results show that charge noise could be the most
important decoherence channel for exchange-coupled spin qubits, and further
development in device design and fabrication is needed to reduce the
sensitivity of exchange coupling to charge fluctuations.  In particular, the
linear scaling of charge fluctuation induced spin dephasing of the exchange
gate architecture with the number of gates is a rather serious problem that
could limit the scalability of exchange-based spin quantum computation.  Our
finding of the charge fluctuation induced spin qubit dephasing time being in a
wide range from ns to $\mu$s points to the need to optimize the double dot
design in order to de-sensitize exchange coupling $J$ to the environmental
charge noise.

\begin{acknowledgments}
We thank J. Petta and C.M. Marcus for useful discussions.  This work was
supported in part by LPS, NSA, and ARO.  
\end{acknowledgments}


\begin{thebibliography}{99}

\bibitem{LD} D. Loss and D.P. DiVincenzo, Phys. Rev. A {\bf 57}, 120 (1998).

\bibitem{Kane} B.E. Kane, Nature {\bf 393}, 133 (1998).

\bibitem{DiVincenzo} D.P. DiVincenzo {\it et al.}, Nature {\bf 408}, 339
(2000).

\bibitem{HD} X. Hu and S. Das Sarma, Phys. Rev. A {\bf 61}, 062301 (2000).

\bibitem{SQCReviews} X. Hu and S. Das Sarma, Phys. Stat. Sol. (b) {\bf 238},
360 (2003); X. Hu, cond-mat/0411012; S. Das Sarma {\it et al.}, Solid State
Commun. {\bf 133}, 737 (2004).

\bibitem{Elzerman} J.M. Elzerman {\it et al.}, Nature {\bf 430}, 431 (2004).

\bibitem{Johnson} A.C. Johnson {\it et al.}, Nature {\bf 435}, 925
(2005); F. Koppens {\it et al.}, Science {\bf 309}, 1346 (2005).

\bibitem{Hatano} T. Hatano, M. Stopa, and S. Tarucha, Science {\bf 309}, 268
(2005).

\bibitem{Petta} J.R. Petta {\it et al.}, Science {\bf 309}, 2180 (2005).

\bibitem{Feher} G. Feher, Phys. Rev. {\bf 114}, 1219 (1959); D.K. Wilson and
G. Feher, {\it ibid.}, {\bf 124}, 1068 (1961).

\bibitem{Fujisawa} T. Fujisawa {\it et al.}, Nature {\bf 419}, 278 (2002); R.
Hanson {\it et al.}, Phys. Rev. Lett. {\bf 91}, 196802 (2003); {\it ibid.}
{\bf 94}, 196802 (2005).

\bibitem{Golovach} V.N. Golovach, A. Khaetskii, and D. Loss, Phys. Rev. Lett.
{\bf 93}, 016601 (2004).

\bibitem{Rogerio} R. de Sousa and S. Das Sarma, Phys. Rev. B {\bf 67}, 033301
(2003); {\it ibid.} {\bf 68}, 115322 (2003).

\bibitem{Hayashi} T. Hayashi {\it et al.}, Phys. Rev. Lett. {\bf 91}, 226804
(2003); J. Petta {\it et al.}, {\it ibid.} {\bf 93}, 186802 (2004).

\bibitem{Astafiev} O. Astafiev {\it et al.}, Phys. Rev. Lett. {\bf 93}, 267007
(2004).

\bibitem{Burkard} G. Burkard, D. Loss, and D.P. DiVincenzo, Phys. Rev. B {\bf
59}, 2070 (1999).

\bibitem{Jung} S.W. Jung, T. Fujisawa, Y. Hirayama, and Y.H. Jeong, Appl.
Phys. Lett. {\bf 85}, 768 (2004); M. Pioro-Ladri\'{e}re {\it et al.},
Phys. Rev. B {\bf 72}, 115331 (2005).

\bibitem{footnote-Barrier} A local barrier potential that does not affect
electron orbitals could lead to quantitative changes to our results, but not
the range of values for the exchange splitting $J$.

\bibitem{growth_direction} The growth direction for the gated quantum dots,
with much higher confinement energy, is regarded as immune to charge
fluctuations and therefore inactive for our considerations.  

\bibitem{Herring} For well separated quantum dots with small overlap,
Heitler-London approximation works quite well.  See C.
Herring, in {\em Magnetism} vol. IIB, ed. by G.T. Rado
and H. Suhl (Academic Press, New York, 1966).

\bibitem{Ono} K. Ono {\it et al.}, Science {\bf 297}, 1313 (2004).

\bibitem{Thorwart} M. Thorwart and P. H\"{a}nggi, Phys. Rev. A {\bf 65},
012309 (2001).

\bibitem{Digital} M. Friesen, R. Joynt, and M.A. Eriksson, Appl. Phys. Lett.
{\bf 81}, 4619 (2002).

\end{thebibliography}
\end{document}